\documentclass[manuscript]{acmart} 
\AtBeginDocument{%
  \providecommand\BibTeX{{%
    \normalfont B\kern-0.5em{\scshape i\kern-0.25em b}\kern-0.8em\TeX}}}

\copyrightyear{2023}
\acmYear{2023}
\setcopyright{acmlicensed}\acmConference[RecSys '23]{Seventeenth ACM Conference on Recommender Systems}{September 18--22, 2023}{Singapore, Singapore}
\acmBooktitle{Seventeenth ACM Conference on Recommender Systems (RecSys '23), September 18--22, 2023, Singapore, Singapore}
\acmPrice{15.00}
\acmDOI{10.1145/3604915.3608856}
\acmISBN{979-8-4007-0241-9/23/09}

\usepackage{acronym}
\usepackage{bbm}

\usepackage{amsmath,amssymb}

\begin{document}

\title{Ex2Vec: Characterizing Users and Items from the Mere Exposure Effect}

\author{Bruno Sguerra}
\authornote{Contact author: \href{research@deezer.com}{research@deezer.com}}
\affiliation{
  \institution{Deezer Research}
    \city{}
  \country{France}
}

\author{Viet-Anh Tran}
\affiliation{
  \institution{Deezer Research}
    \city{}
  \country{France}
}

\author{Romain Hennequin}
\affiliation{
  \institution{Deezer Research}
    \city{}
  \country{France}
}

\renewcommand{\shortauthors}{Sguerra et al.}

\begin{abstract}
The traditional recommendation framework seeks to connect user and content, by finding the best match possible based on users past interaction. However, a good content recommendation is not necessarily similar to what the user has chosen in the past. As humans, users naturally evolve, learn, forget, get bored, they change their perspective of the world and in consequence, of the recommendable content. One well  known mechanism that affects user interest is the Mere Exposure Effect: when repeatedly exposed to stimuli, users' interest tends to rise with the initial exposures, reaching a peak, and gradually decreasing thereafter, resulting in an inverted-U shape. Since previous research has shown that the magnitude of the effect depends on a number of interesting factors such as stimulus complexity and familiarity, leveraging this effect is a way to not only improve repeated recommendation but to gain a more in-depth understanding of both users and stimuli.  In this work we present~\ac{Ex2Vec} our model that leverages the Mere Exposure Effect in repeat consumption to derive user and item characterization and track user interest evolution. We validate our model through predicting future music consumption based on repetition and discuss its implications for recommendation scenarios where repetition is common.
\end{abstract}



\maketitle

\section{Introduction}

\textit{``No man ever steps in the same river twice, for it is not the same river and he is not the same man''} is a quote credited to Heraclitus, the pre-Socratic Greek philosopher, six hundred years before Christ. Heraclitus' celebrated phrase illustrates his view of an ever dynamic world and individual, both in a perpetual state of change. Heraclitus understood something that is often overlooked in recommender systems literature, i.e., the past rarely repeats itself. The traditional recommendation framework seeks to connect user and content, by finding the best match possible based on users past interaction~\cite{koren2009matrix,rendle2021item}. However, a good content recommendation is not necessarily similar to what the user has chosen in the past. One limitation of basing future interaction on what happened in the past is that it ignores the fact that both sides of the problems are dynamic. As humans, users naturally evolve, learn, forget, get bored, they change their perspective of the world and in consequence, of the recommendable content. The development of effective recommender systems, therefore, requires researchers and practitioners to account for this dynamism.  

The last years have witnessed an increase in the number of works that attempt to embed recommender systems with psychology-derived knowledge on human behavior~\cite{lex2021psychology, curmei2022towards, sguerra2022discovery, reiter2021predicting}. There is an extensive body of psychological research concerned with different effects that affect users and their preferences when dealing with content consumption~\cite{curmei2022towards}. One example of a well studied effect that has implications for recommender systems is the \ac{MEE}. The effect states that the mere exposure of an individual to a stimulus is enough to result in the development of a positive attitude towards the stimulus~\cite{zajonc1968attitudinal}.  
Therefore, the~\ac{MEE} should play a significant role in shaping the evolution of a user's interest in various feed recommendation scenarios where repeated consumption is prevalent, such as music streaming.

In this article we present~\ac{Ex2Vec}, our model that leverages repeat consumption to learn joint user and item representations. Based on the finding that the magnitude of~\ac{MEE} depends on both user and stimuli~\cite{bornstein1989exposure,sguerra2022discovery}, \ac{Ex2Vec} learns to predict the user's interest evolution from repeated exposure by characterizing both user and item. We believe that the  representation learned by~\ac{Ex2Vec} can not only improve recommendation where repeated consumption is common, but more importantly,  provide researchers with new information about users and stimuli, such as perceived familiarity, complexity and more. 

To summarize, our paper includes three major contributions: (1) an analysis conducted to confirm the existence of the Mere Exposure in music streaming consumption;  (2) \ac{Ex2Vec}, a model that allows for both user and item characterization and the prediction of repetitive consumption behavior; (3) the publication of the collected data to the research community.

\section{Related Work}
\label{sec:related work}

\textbf{Mere exposure effect}. The \ac{MEE} is a long known effect in psychology which states that the ``mere'' exposure of an individual to a given stimulus is enough to increase both recognition and affect (read liking or pleasantness) towards said stimulus~\cite{zajonc1968attitudinal}. In ~\cite{bornstein1989exposure, montoya2017re}  the reader will find a comprehensive meta-analysis on over forty years of controlled experiments involving it. The effect is robust and consistent across different types of stimuli, including images, words, simple sounds, and music.~\cite{zajonc2001mere, bornstein1989exposure}. Previous research has identified several factors that can affect the strength of the \ac{MEE}, such as a person's personality, age, the order and length of presentation, and stimulus complexity~\cite{bornstein1989exposure}. While exposure to stimuli results in increased affect, the effect is not monotonic and eventually reaches a maximum point, after which further repetitions can lead to satiation. The resulting behavior is described by an inverted-U shaped curve, a pattern rather consistent across different domains~\cite{montoya2017re}.  Among the many models attempting to explain the effect~\cite{bornstein1989exposure}, we call attention to Berlyne's two factor model~\cite{berlyne1970novelty}. Where the effect is described as the interaction of two-factors: (1) a rising habituation factor over exposures coupled with (2) a tedious factor. The habituation factor occurs during the initial repeated exposures, as the individual becomes accustomed to the stimuli. This familiarity allows previously imperceptible details or aspects of the stimuli to become accessible to the individual, until eventual satiation or boredom sets in. Bornstein described boredom as a limiting factor of the~\ac{MEE}~\cite{bornstein1989exposure} and is a common aspect of repeat consumption in many domains, where users lose interest after over-consuming an item~\cite{benson2016modeling, kapoor2015just, anderson2014dynamics}. Not only restricted to the~\ac{MEE}, Berlyne provides a more general theory for aesthetic preference~\cite{berlyne1960conflict}, where preference follows an inverted-U shape over a number of collative variables, such as complexity, novelty, familiarity, among others. A theory that has been tested extensively and is rather consistent~\cite{chmiel2017back}.

There are many interesting aspects of the~\ac{MEE} for recommender systems. Curmei et al., in~\cite{curmei2022towards}, describe how the~\ac{MEE}, among other known psychological effects, can create feedback loops changing users' interest. They formulate the mere exposure effect by linearly modifying the user preference vector towards the interacted item. In consequence, when a user interacts with a given content, the vector representing their taste is moved towards the content to represent that ``whenever users are exposed to content, this makes them like this content more''. This has obvious limitations, for instance, the fact that someone is discovering a new interest does not mean they stopped liking their previous preferences: one can like both The Beatles and The Rolling Stones. Moreover, this simple linear model does not account for the saturation of the effect and its characteristic inverted-U shape. In~\cite{sguerra2022discovery} we analyzed the actual consumption patterns of newly released music on Deezer, a music streaming service. We show that, for users listening repeatedly to newly released songs, the interest curve follows an inverted-U shape over exposures, characteristic of the~\ac{MEE}. 

\textbf{Psychology-informed recommender systems}. In recent years, there has been a growing interest in incorporating psychological knowledge into recommender systems, moving away from purely algorithmic models and towards greater interpretability. In their recent survey~\cite{lex2021psychology}, Lex et al. present efforts made along three axes: (1) cognition, (2) personality and (3) affect. In this section we revise the literature of the first category as it is the most relevant to our research. Cognitive-inspired recommender systems exploit mental processes of memory, attention and learning for modeling user behavior and adapting feedback to improve recommendation. Considering memory, in~\cite{reiter2021predicting}, authors introduce a time-based exponential decay factor to weight explicit feedback and improve the accuracy to the collaborative filtering framework. The premise being that users' taste evolution can be modeled as a type of information forgetting, with their time decay factor inspired by Ebbinghaus' forgetting curve~\cite{ebbinghaus1913memory}. The same curve is used in~\cite{hu2011nextone} to model a ``freshness'' factor of a listened song, in order to modulate recommendations and avoid overexposure.  One of the most popular models of human cognition employed in recommender system's literature is Anderson's cognitive architecture ATC-R~\cite{anderson2009can}. This fixed architecture has been used to model a multitude of cognitive tasks and simulate human cognitive performance. Authors in~\cite{reiter2021predicting, kowald2020utilizing, kowald2017temporal} employ  a specific module of ACT-R, the declarative one,  to model either the reuse of hashtags on twitter or the relistening of music.  ACT-R's declarative module represents a ``window to the past'' where learned information or facts can be accessed and serves to model human memory processes. Since it accounts for both frequency and recency, it has been quite successful in modeling repeat consumption as in the work cited above. 

The literature discussed describes various efforts to incorporate knowledge of human cognitive processes and limitations into recommender systems. While psychology offers a wealth of information on the~\ac{MEE}, little research has been conducted on using this knowledge to enhance recommender systems. In the upcoming sessions, we  aim to fill this gap by exploring its potential. 

\section{The Mere Exposure Effect in Music Consumption}

Among the many application domains where recommender systems are used, we posit that music streaming is one where the~\ac{MEE} is more easily observed and perhaps useful to account for. First, the engagement of listening to a music track is quite light when compared to watching movies or buying products, both in terms of time and money, increasing the overall number of user interactions~\cite{schedl2019deep}. Additionally, repetition is a rather common phenomenon in music consumption, allowing for tracking the evolution of interest based on exposure or repetition~\cite{sguerra2022discovery}, while for other  domains such as books and movies, repetition behavior emerges at a higher level of abstraction~\cite{kapoor2015just}. Also, the number of items in commercial music catalogs has a magnitude of tens of millions of tracks that is quite diverse in terms of popularity, language and others, that can be easily controlled for exposure. Lastly, music itself is a complex stimuli that has been studied in a large body of literature of the~\ac{MEE}~\cite{chmiel2017back,szpunar2004liking}. For these reasons, in this article we focus on the domain of music consumption in streaming platforms.

In previous work we focused on newly released albums to study the~\ac{MEE} in music consumption~\cite{sguerra2022discovery}. We show that for the new albums, the listening probability follows an inverted-U shape over exposures, while for classic rock music, the probability of listening decreases monotonically with exposures. Accordingly,  accounting for newly released songs reduces the chances of the music being known by the users and that the rise of interest corresponds to a form of ``learning'' as users get habituated to the song. Therefore we also focus our research on newly released music tracks. In the next section we outline some characteristics of the~\ac{MEE}.

\subsection{Data}
\label{sec:data}
The present study utilizes song listening histories obtained from Deezer, a well-established music streaming platform. We isolated the listening history, from August to December 2022, related to the tracks released during the month of August 2022. We filter out tracks that were not listened to by at least 20 different users and users that did not listen to at least 20 different tracks. To increase the likelihood of the user's interest evolution having finished its course, we remove from the data the entire user-item consumption sequences that appear after 80\% of the considered time window.  Every row of the resulting dataset contains an user identifier, an item identifier, a timestamp and the user listening time.

\subsection{Empirical Observation of the MEE }
\label{sec:empirical}
\begin{figure*}
    \centering
    \includegraphics[width=0.5\linewidth]{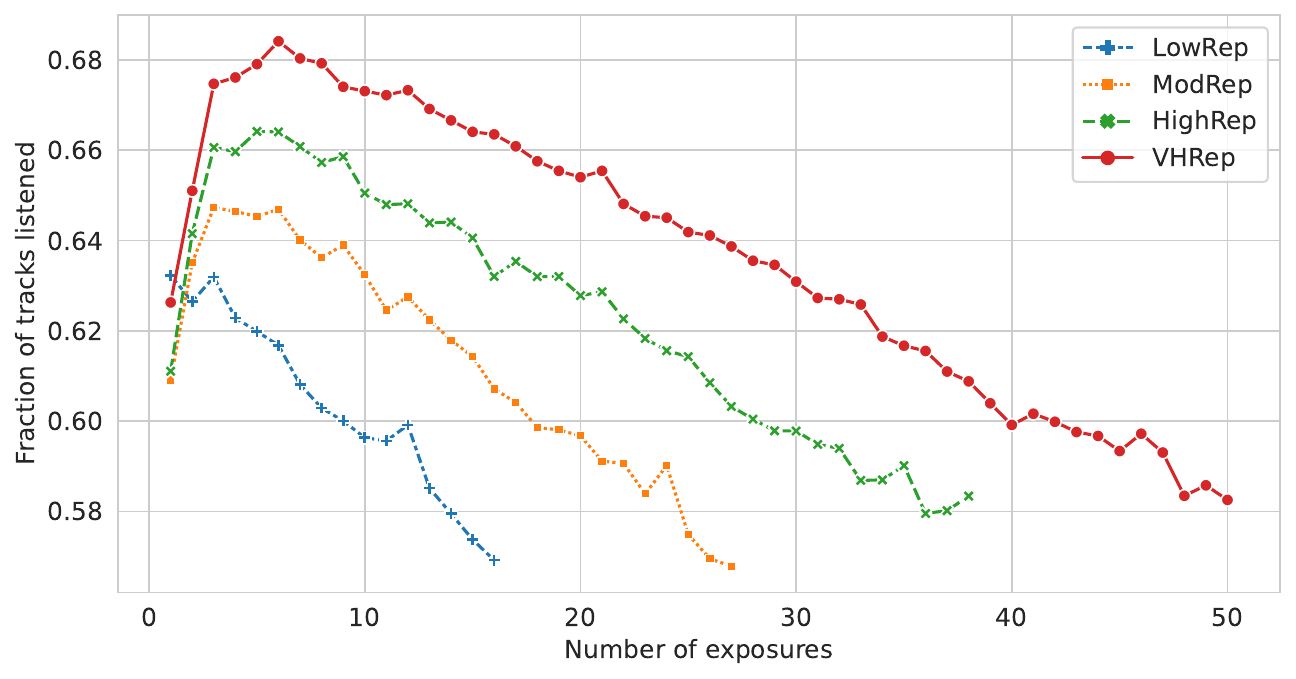}
    \caption{The fraction of listening events per exposure for each class of repetition.}
\label{fig:macro}
\end{figure*}

In this section we check to see if the \ac{MEE} is observable in the data.  For accessing the evolution of interest over exposures, the number of exposures is of most importance. Here we consider only consumption sequences with 5 or more exposures. To obtain a more representative view of the different behaviors related to the number of exposures, we sampled the data based on the total size of the consumption sequence. We filtered out the entire user-item consumption sequences with more than 50 exposures (less than 0.05 of the data) and discretized the dataset into 4 classes with the same repetition interval based on the total number of repetitions of a pair user-item: \textbf{(i) LowRep}, the lowest total number of repetitions from 5.0 to 16.0, with $|U| = 29.6$k distinct users and $|I| = 9.2$k distinct tracks; \textbf{(ii) ModRep}, number of repetitions from 17.0 to 27.0, $|U| = 26$k and $|I| = 7.6$k; \textbf{(iii) HighRep}, high number of repetitions from 28.0 to 38.0, $|U| = 22.5$k and $|I| = 6.9$k and \textbf{(iv) VHRep}, the highest total number of repetitions, from 39 to 50, $|U| = 19.8$k distinct users and $|I| = 6.5$k distinct items.  The resulting dataset comprehends the consumption history of $|U| = 52$k and $|I| = 13.7$k and  contains about  4.7M lines. Figure~\ref{fig:macro} depicts the fraction of the tracks that were listened to at a given exposition $j$. Here we consider that the user listened to an item, $L = 1$, if the user listens to more than 80\% of the duration of the track and $L = 0$ otherwise. At first exposure, all classes had a similar listening ratio ($\approx0.62$), subsequently evolving differently. If this fraction is taken as a proxy for the user's interest, the maximal attained value increases with the total number of exposures, later on decreasing at a similar ratio. All classes (with the exception of LowRep where interest decreases with exposures) exhibited an inverted-U shape, which is a characteristic of the \ac{MEE}.

\begin{figure*}
    \centering
    \includegraphics[width=0.7\linewidth]{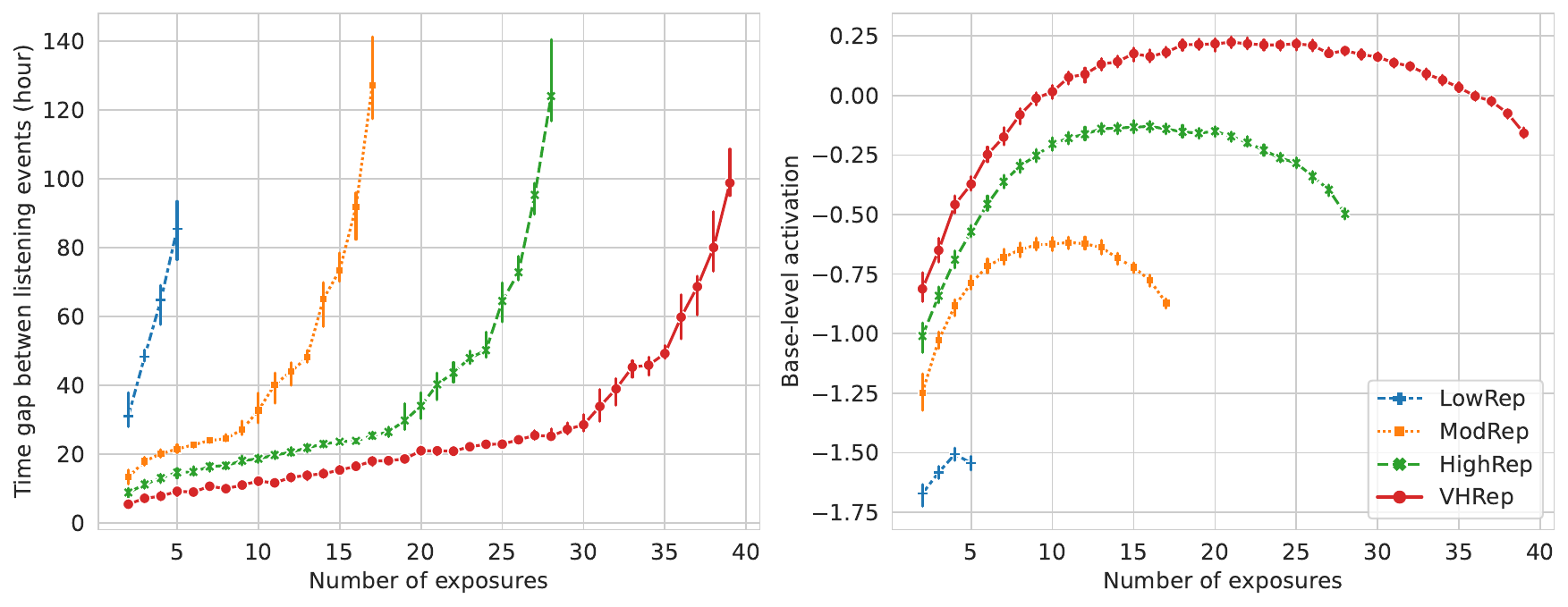}
    \caption{Left: Median time gap between listening events with 0.95 confidence interval over the number of exposures for each class of repetitive behavior. Right: The median value of the base-level activation with 0.95 confidence interval over the number of exposures for each class.}
\label{fig:base_level}
\end{figure*}

Although the fraction of listening events can be used as a proxy of interest, it ignores the frequency of consumption.  As previously discussed, presentation sequence is one of the modulating factors of the effect~\cite{bornstein1989exposure} . In order to account for the presentation sequence frequency, we employ the popular ATC-R's declarative module. Part of the declarative module is the base-level activation that accounts for repetition, frequency and recency and is given by: $B_i(t) = \ln \sum_{j=1}^{n} \mathbbm{1}_{t>t_j}(t-t_j)^{-d} $, where $n$ is the number of access to a given information $i$ (in our case, the number of past consumption of song $i$), $t_j$ is the time of the $j$-th access to that information (in our case, time of the $j$-th consumption of song $i$), $\mathbbm{1}_{t>t_j}$  is the indicator function that is $1$ when $t>t_j$ and $0$ otherwise, and $d$ is a decay parameter ($d>0$).  We compute the base-level activation for the consumption sequences in our dataset (only considering a repetition when $L = 1$) as it provides a more accurate representation of interest: if users are interested in a song, not only they will repeat it more, but with increased frequency. Here we make a distinction between exposure and repetition, where repetitions are exposures where the user actually consumed the item, or in this case, when $L=1 $. Following ACT-R's community we set $d  = 0.5$, which is a value that has surfaced as the default setting for this parameter~\cite{anderson2009can}. Figure~\ref{fig:base_level} depicts on the left side, the evolution of the median time gap in hours between repetitions at a given exposure (note that the time gap at repetition $j$  is given by $gap_j = timestamp_j - timestamp_{j-1}$) and on the right the evolution of the median base-level activation value per exposure (note that in the figure, we only present the values after the first consumption). We consider the most popular repetitions for each class: LowRep (5), ModRep (17), HighRep (28), and VHRep (39). The left side of Figure~\ref{fig:base_level} shows periods of relatively stable interest (except for the LowRep class) before slowly growing, indicating satiation. The observed behavior for the base-level activation is also consistent with the \ac{MEE} theory, where users tend to consume a song more frequently after initial exposure then slowly losing interest.

\section{(Mere) Exposure 2 Vec}

The last section was concerned with observing the \ac{MEE} in consumption of newly released songs. In this section, we define  the \ac{Ex2Vec} model and describe how it is able to characterize both users and items based on the dynamics of the ~\ac{MEE} as well as track users' interest evolution over repetitions.  Subsection~\ref{sec:dynamics} reviews the concepts behind \ac{Ex2Vec} and Subsection~\ref{sec:model} describes it.

\subsection{MEE Dynamics}\label{sec:dynamics}

Berlyne's theory of aesthetic preference posits that preferences follow an inverted-U shape as a function of collative variables, such as uncertainty, familiarity, and novelty~\cite{berlyne1960conflict,berlyne1970novelty,chmiel2017back}. Accordingly, interest for a stimulus is expected to start low, increase with familiarity, reach a peak, and then decrease, as observed in Section~\ref{sec:empirical}.  Uncertainty has also been related to musical pleasure as part of the enjoyment derives from being able to anticipate some aspects of the song while still being surprised by others~\cite{cheung2019uncertainty, gold2019predictability}. However, too little or too much uncertainty can reduce the pleasure of the experience. Repeating the same song therefore would increase familiarity, and eventually decrease uncertainty, resulting in the inverted-U shape. We relate this increase of familiarity and decrease of uncertainty as a learning process and to model this mechanism we employ ACT-R's declarative module. 

Originally, ACT-R's declarative module was proposed to model the dynamics of human memory. The base-level activation, presented above, is part of a series of activation energies that modulate how easily and readily an information is retrievable in memory.  In ACT-R, the probability of a memory $i$ being retrieved is given by $P_i = \sigma(\frac{A_i - \tau}{s})$, where $\sigma(x) = 1/(1+e^{-x})$ is the sigmoid function; $\tau$ is the activation threshold (below which the odds of the information being retrieved are low); $s$ is a smoothness parameter; and $A_i$ is the activation energy of information $i$,  given by: $
A_i =  \sum_{j \in C} W_jS_{ji} + B_i$, where $B_i$ is the same base-level activation already presented, and the other component represents a sort of similarity of information $i$  with the previously stored knowledge (for more details we recommend~\cite{anderson2009can}). The intuition being that memory recall depends on two factors, (1) repetition and (2) past knowledge. However, with time, memories tend to degrade given the decay factor $d$. This activation energy $A_i$ provides us with a proxy of how well encoded in memory a given piece of information is (or in our case a given song) and how easily a given piece of information can be accessed or recalled. It is worth noting that not only liking, but also memory and recognition, have been studied as a function of exposure and may be interconnected~\cite{szpunar2004liking}. 

Following Berlyne's theory of aesthetic pleasure, interest should then follow an inverted-U shaped pattern along the evolution of $A_i$, if the $A_i$ is too small, the stimuli might be too uncertain/complex to the user, with repetition, as the user gains a better understanding of the song, interest might develop until overexposure where it falls back again.  These are the two mechanisms modeled by~\ac{Ex2Vec} that are formalized in the next section.

\subsection{Ex2Vec Definition}
\label{sec:model}

Much as traditional collaborative filtering techniques such as Matrix Factorization, \ac{Ex2Vec} projects users and items into a latent space. Users are described by an embedding vector noted $\mathbf{u}$ and items by an embedding vector $\mathbf{v}$, both with the same number of latent dimensions $D$. We assume the distance $d(\mathbf{u},\mathbf{v})$ to behave as a number of collative variables such as perceived uncertainty/complexity etc. As the user repeats a song, we linearly modify this distance similarly to~\cite{curmei2022towards}. However, instead of moving the user embedding towards the item at every new interaction, \ac{Ex2Vec} modulates only the relative distance between $\mathbf{u}$ in respect to item $i$. This approach better reflects a user's tastes, as discovering new items does not necessarily mean losing interest in previous preferences. When $d(\mathbf{u},\mathbf{v})$ diminishes it indicates that the user has gained a better understanding of the item, representing a form of learning. Since humans forget with time, \ac{Ex2Vec} accounts for a time-decay factor based on ACT-R's declarative module. If the discovery of new items is seen as a form of learning, it is natural to base the learning evolution in the dynamics of human memory provided by the declarative module. At time $t$ for item $i$, the distance  $d_{u,i}(t)$ is given by:

\begin{equation}
    d_{u,i}(t) = max(d(\mathbf{u},\mathbf{v}) - \lambda_u \sum_{j = 1}^{n} \mathbbm{1}_{t>t_j}(t-t_j + c)^{-d}, 0),
\label{eq:user_translation}
\end{equation}
where $n$ is the number of past consumption of item $i$, $t_j$ stands for the time of past consumption $j$ of item $i$ ($t_j < t$); $d$ is the decay parameter similar to ACT-R's declarative module and $\lambda_u$ > 0 is the step size, regulating how much to change the distance. We make the $\lambda_u$ a summation of a global $\lambda$ and a user specific bias: $\lambda_u = \lambda + \lambda_{u}^b$. Note that we removed the $\ln$ transformation from ACT-R's base-level to ensure the minimum value to subtract from the distance to be zero and we added a cutoff term $c > 0$ to keep the change in distance bounded when $t - t_j$ is too small. Moreover, we take the maximum value between 0 and the modified distance in order to, at the very limit, have the distance $d_{u,i} = 0$. \ac{Ex2Vec} models the \ac{MEE} as a form of learning. The base distance  $d(\mathbf{u},\mathbf{v})$, similar to $A_i$ from ACT-R's declarative module, models the user's base familiarity/knowledge with a given item. The more the user repeats a given song, the smaller $d_{u,i}(t)$ will be. With time, without new repetitions, $d_{u,i}(t)$ will return to  the base distance $d(\mathbf{u},\mathbf{v})$, indicating the forgetting of the item. The base activation, given by the term $\sum_{j = 1}^{n} \mathbbm{1}_{t>t_j}(t-t_j + c)^{-d}$
 modulates the distance evolution.

In order to account for the inverted-U shape characteristic of the~\ac{MEE}, instead of simply using the distance  $d_{u,i}(t)$ as a proxy for the user's interest, we introduce a quadratic term for the user's interest in an item at time $t$:

\begin{equation}
    I(\mathbf{v},\mathbf{u},t) = \alpha d_{u,i}(t) + \beta d_{u,i}(t)^2 + \gamma + b,
\label{eq:quadratic}
\end{equation}
Where $\alpha$, $\beta$ and $\gamma$, are global parameters of the quadratic function and $b$ is the sum of a user and item bias. The main functioning of ~\ac{Ex2Vec}, inspired by the dynamics of the mere exposure effect, are described by equation~\ref{eq:user_translation} and equation~\ref{eq:quadratic}.

\section{Learning User and Item characterization from MEE}

In order to learn the user and item characterizations from data, we take inspiration from NeuMF~\cite{he2017neural}. Much as in NeuMF, as input values we have a binarized sparse identity vector for users and items followed by a fully connected embedding layer, that projects both binary vectors into the dense ones $\mathbf{u}$ and $\mathbf{v}$ with dimension $D$. From the embedding vectors we compute the base distance $d(\mathbf{u},\mathbf{v})$ (we used the Euclidean distance as it provides the best performance), modulating it  in accord to the  factor $\lambda_u(\sum_{j = 1}^{n} \mathbbm{1}_{t>t_j}(t-t_j + c)^{-d})$ of equation~\ref{eq:user_translation} obtaining the final distance  $d_{u,i}(t)$.  Lastly we apply the quadratic function of equation~\ref{eq:quadratic} to the computed distance obtaining the user interest that is used as input of a sigmoid function obtaining the final predicted score $y'_{u,i}$  constricted to a range of $[0,1]$. Since we are modeling the listening behavior of users, i.e., if a user listens to a song or not, we train our model with both instances when $L = 1$ or $L = 0$. Much as in~\cite{he2017neural} we use the log loss for training with a $L2$ regularization term. We employ the Adam algorithm~\cite{kingma2014adam}, jointly learning both the user item representations and the parameters: $c$, $\lambda$, $\lambda_{b}^u$, $b$ and $\alpha$, $\beta$ and $\gamma$ of the quadratic function. We fix the decay parameter to $-0.5$ according to ACT-R's literature. 

\subsection{Evaluation}

Since \ac{Ex2Vec} predicts interest evolution with repetitions. We evaluate~\ac{Ex2Vec} performance in predicting the consumption sequences i.e., if a user at a given exposure is going to listen to a given song ($L=1$) or not ($L=0$). This is a task consistent with the real case scenario of creating playlists for music discovery or of new released tracks. In the real case scenario, by predicting the interest of users over time,  given the number of past consumption, songs can be added or removed dynamically to these playlists, removing songs that were already overexposed or keeping items whose interest peak takes longer to attain. We train \ac{Ex2Vec} on the same listening histories presented in Section~\ref{sec:data}. To ensure that every user and song have enough interactions to be properly characterized, such as in~\cite{sun2020we}, we implement a $k^{item}$ and $k^{user}$ pre-processing step. This step  consists in recursively filtering the data until all users have interacted with at least  $k^{item}$ items and every item was consumed by at least  $k^{user}$ users. We empirically decided on $k^{item} = k^{user} = 30$, which results in a dataset of about 1.5M interactions, $|U|=3.6$k and $|I| = 878$. For the validation and test set, we randomly selected four items for each user in the dataset and entirely removed from the training set any instances of the user being exposed to these items. While the items themselves are still included in the training set, any interaction of a user with their sampled items were removed to prevent them from influencing the training process. Two items per user of the removed instances are used as validation set and the other two as test set. We implement~\ac{Ex2Vec} with pytorch and make both the code and data available online\footnote{https://github.com/deezer/ex2vec}. 

We compared our model with four baselines for predicting repeated behavior: (1) $SLRC$, a sequential recommendation model that learns item-specific temporal patterns of re-consumption (with the base intensity given by Bayesian Personalized Ranking)\cite{wang2019modeling}. (2) $BL$, which uses ACT-R's base-level activation as a proxy for interest. (3) $BL_{fit}$, which utilizes the base-level activation with a fitted decay parameter, as demonstrated in~\cite{kowald2017temporal} to be effective in predicting relistening events. (4) $Prev$ simply assumes that a user will relisten at the next exposure if they have listened to it in the previous exposure. Although simple, with this baseline, we intend to shift the inverted-U shape by a single exposure. For predicting the relistening of a song, for~\ac{Ex2Vec}, $SLRC$, $BL$, and $BL_{fit}$, we discretized two classes: $L=1$ or $L=0$ based on: the interest (for \ac{Ex2Vec}), the intensity for $SLRC$, and the two base-level values. The discretization threshold was defined as the value that maximizes the balanced accuracy on the validation set. Both~\ac{Ex2Vec} and $SLRC$ were set with the same embedding dimension ($D = 64$) and optimized on the validation set with learning rate values in \{5e-5, 0.0002, 0.00075, 0.001\} for 100 epochs. For~\ac{Ex2Vec} we initialized the parameters as following: $\alpha = 1.0$, $\beta = -0.0065$, $\gamma = 0.5$, $\sigma = 1.0$, $c = 3.0$, the user and item embeddings where initialized from $\mathcal{N}(0,1)$. \ac{Ex2Vec} shows an overall better performance as shown in Table~\ref{table: evaluation}.

\begin{table}
\centering
\resizebox{0.6\columnwidth}{!}{
\begin{tabular}[t]{ccc}
\hline
  Model  & Balanced Accuracy (in \%)&  Weighted F1-score (in \%) \\
\hline
$Ex2Vec$ & \textbf{64.27 $\pm$ 0.4} &  \textbf{64.3 $\pm$ 0.38} \\
$SLRC$ & 58.16 $\pm $0.16 & 58.15 $\pm $0.17\\
$BL$ & 57.81 $\pm $0.28 & 58.13 $\pm$ 0.31 \\
$BL_{fit}$ & 57.48 $\pm$ 0.35 & 57.9 $\pm$ 0.42 \\
$Prev$ & 59.88 $\pm$ 0.32 & 59.83 $\pm$ 0.32 \\
\hline
\end{tabular}}
\caption{Experimental results. Scores averages computed on five splits of the test set.}
\label{table: evaluation}
\end{table}

\subsection{Further Analysis}

\begin{figure*}[h]
    \centering
    \includegraphics[width=0.5\linewidth]{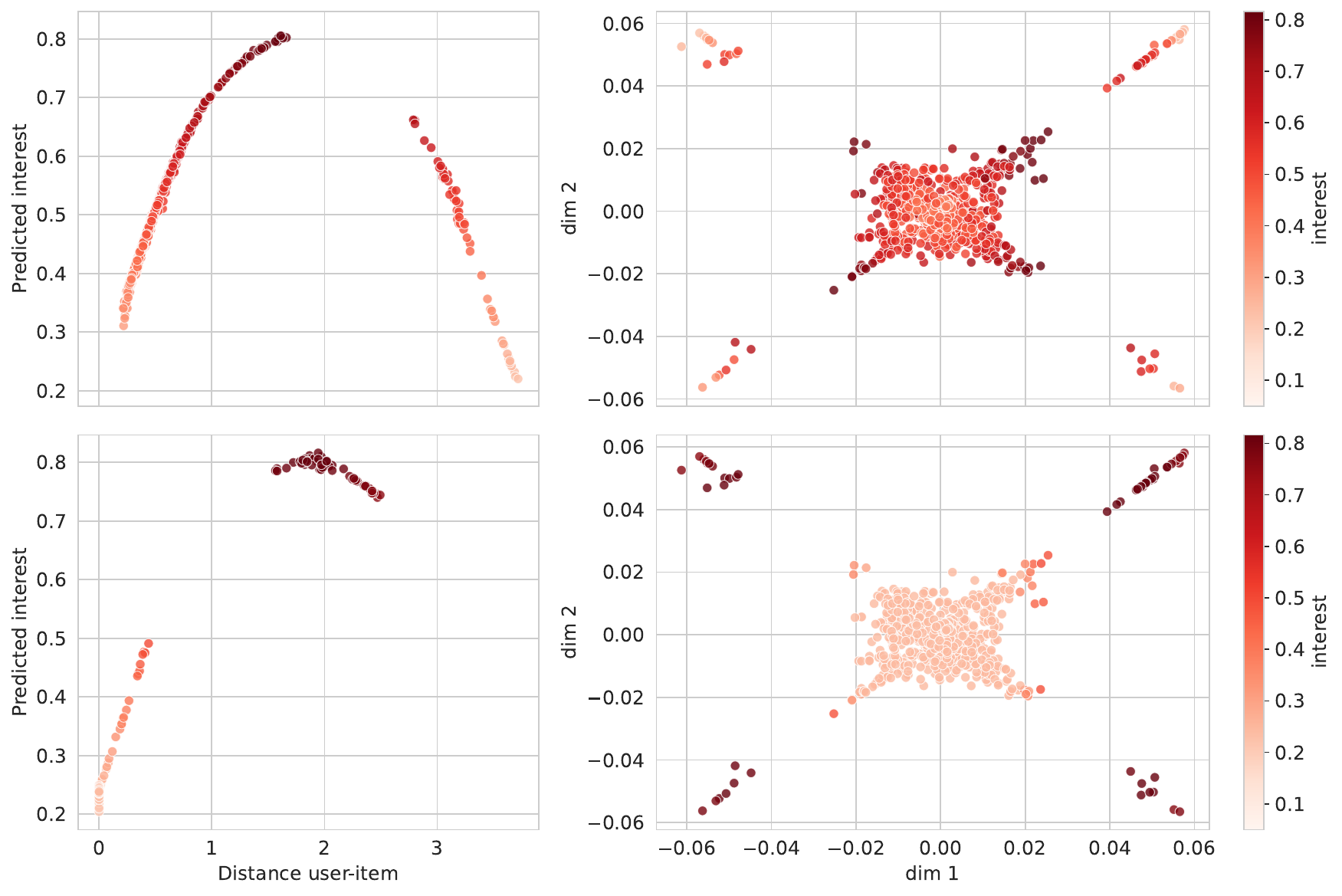}
    \caption{Above left: the interest over the computed $d_{u,i}(t)$ distance with no base-level activation; above right: the first and second dimensions of the item embeddings with the corresponding predicted interest with no base-level activation; below left: the predicted interest given the distance with a higher base-level activation value; below right: the first and second dimensions of the item embeddings with the higher base-level computed interest.}
\label{fig:embeds}
\end{figure*}

However, more than the predictive performance of \ac{Ex2Vec}, the main interest comes from the user/item representations learned. Instead of moving users and items closer in space to reflect interest, as traditional collaborative filtering techniques, \ac{Ex2Vec} will position them based on repetitive behavior and the quadratic interest function. Therefore, items too close to a user have low predicted interest, as for items too far from the user, according to the inverted-U shape. For illustration, on top of Figure~\ref{fig:embeds} depict, on the left,  the predicted interest given the computed $d_{u,i}(t)$ distance between the items and a sampled user with no base-level activation, i.e. before any repetition. On the top right the first and second dimensions of the learned item embeddings are depicted with the corresponding predicted interest. Much as Berlyne’s collative variables, interest is following an inverted-U shape along the distance: items that are closer and furthest from the user have smaller predicted interest. The items with the optimal predicted interest are thus found in a ring-like shape around a user reference. When the user starts consuming these items the distance user-item diminishes according to the base-level activation. The bottom of Figure~\ref{fig:embeds} depicts the same items as above but with the interest computed with more frequent or recent repetition  (base-level activation = 11.12).  The more a user listens to the songs, the smaller the distance and the interest therefore either decreases or increases accordingly to the inverted-U shape based on the relative position of user and item. Therefore, the distances in the embedding space serve as a proxy of how much repetition a user needs in order to either start to like a new song or start to lose interest in another. This allows for recommendation to be made by balancing items that have high predicted interest, stop recommending items that are saturated or even repeatedly recommending novel items that are further away and take more consumption for the user to start liking them. Balancing these cases might be a way to ease the cognitive charge of discovering different content and avoid boredom in feed-like applications for instance.

\section{Conclusion and Future Directions}
We introduce~\ac{Ex2Vec}, a model for leveraging repeated exposure to characterize users and items. We demonstrate that \ac{Ex2Vec} has a dual capability, allowing for both user and item characterization and prediction of relistening behavior. We posit that~\ac{Ex2Vec} enables new recommendation paradigms by tracking the user's interest and balancing recommendation  at different levels of familiarity, promoting learning and habituation while preventing boredom.  Furthermore, as \ac{Ex2Vec} positions users and items based on the evolution of user interest through repetition, it is likely to reflect some of Berlyne's collative variables, such as uncertainty and familiarity, among others. We believe that \ac{Ex2Vec} serves as an example of how leveraging psychological research allows not only for the improvement of recommender systems, but to help leveraging the rich human behavior data available to understand users and stimuli better. 

One limitation of Ex2Vec is its treatment of relative distances between users and items, which results in changes of interest only related to the interacted item. To address this, future enhancements should explore modeling the relational dynamics among items, accounting for the effects of listening to closely related items on users' learning and forgetting processes. Additionally, fluctuations in users' attention, context, and intention, which often lead to significant changes in preferences and learning abilities, should also be accounted for in the future. Finally, future work should investigate the relationship between the learned embeddings and Berlyne's collative variables in more depth. If this relationship holds, \ac{Ex2Vec} could become a valuable tool for inferring stimuli characteristics, such as relative familiarity and complexity, as well as user-specific traits, such as personality and knowledge.

\acrodef{MEE}[MEE]{Mere Exposure Effect}
\acrodef{LE}[LE]{Listening Event}
\acrodef{MUIR}[MUIR]{Mere-exposure-based User-Item Representation}
\acrodef{Ex2Vec}[Ex2Vec]{(Mere) Exposure2Vec}

\bibliographystyle{ACM-Reference-Format}
\bibliography{discovery}
\end{document}